\newif\ifpdf
\ifx\pdfoutput\undefined
  \pdffalse 
\else
  \pdfoutput=1 
  \pdftrue
\fi

\documentclass[slac_one]{revtex4}
\ifpdf
  \usepackage[pdftex]{graphicx}
  \usepackage[pdftex,dvipsnames,usenames]{color}
  \DeclareGraphicsExtensions{.jpg, .pdf, .tif, .eps}
\else
  \usepackage[pdftex]{graphicx}
  \DeclareGraphicsExtensions{.eps}
\fi
\usepackage{fancyhdr}
\newcommand{\MET}{E_T\hspace{-2.8ex}/\hspace{1.6ex}}
\newcommand{\squ}{\tilde{q}}
\newcommand{\sbo}{\tilde{b}}
\newcommand{\sto}{\tilde{t}}
\newcommand{\glu}{\tilde{g}}
\newcommand{\cha}{\tilde{\chi}^{0}}
\newcommand{\neu}{\tilde{\chi}^{+}}

\begin{document}

\title{Inclusive Search for Squarks and Gluinos Production and Search 
for Sbottom from Gluino Decay at CDF } 

%

\author{G. De Lorenzo$^{1}$, M. D'Onofrio$^{1}$, O. Gonzalez-Lopez$^{2}$, M. Martinez-Perez$^{1}$, M. Vidal$^{2}$.}
\affiliation{$^{1}$IFAE/ICREA, Barcelona, Spain}
\author{\small{\textit{$^{2}$CIEMAT, Madrid, Spain}}}

\begin{abstract}
We present preliminary results on a search for squarks and gluinos in proton-antiproton collisions with a center-of-mass energy of 1.96 TeV and based on about $2.0\;fb^{-1}$ of data collected by the CDF detector in the Tevatron Run II. 
Events with multiple jets of hadrons and large missing transverse energy in the final state are studied within the framework of minimal supergravity (mSUGRA) and assuming R-parity conservation. 
The results are compared to Standard Model (SM) predictions, and limits on gluino and squark masses are extracted. 
A specific search for the supersymmetric partner of the bottom quark produced from gluino decays is carried out using a sample of events with missing transverse energy and two or more jets in the final state where at at least one of the two leading jets is b-tagged. 
Good agreement is found between data and SM predictions, and limits on gluino and sbottom masses are extracted.
\end{abstract}

\maketitle

\thispagestyle{fancy}


\section{INCLUSIVE SEARCH FOR GLUINO AND SQUARK PRODUCTION} 

Supersymmetry (SUSY)~\cite{SUSY} is regarded as one of the most promising fundamental theories to describe
physics at arbitrary high energy beyond the Standard Model.
In mSUGRA~\cite{mSUGRA} symmetry breaking is achieved via gravitational interactions and
the vast SUSY parameter space is reduced to only five parameters: 
the common trilinear coupling at the GUT scale $A_0$; the sign of the Higgsino mixing parameter $sign(\mu)$; the ratio of the Higgs vacuum expectation values tan$\beta$; the common scalar and gaugino masses at the GUT scale $m_0$ and $m_{1/2}$, respectively.
At the Tevatron, the production of squarks ($\tilde{q}$) and gluinos ($\tilde{g}$), super-partners of quarks and gluons, constitutes one of the most promising channels because of the strong couplings involved.  
The cascade decay of gluinos and squarks into quarks and gluons will result in a final state consisting of several jets plus missing transverse energy ($\MET$) coming from the neutralinos, which leave CDF undetected.  

The inclusive search~\cite{GLUSQUA} for $\squ$ and $\glu$ is performed using 2.0 $fb^{-1}$ of data collected by the CDF detector in Run II~\cite{CDF}.  
A mSUGRA scenario with $A_0=0$, $\mu<0$ and tan$\beta=5$ is assumed.  
The gluino-squark mass plane is scanned via variation of the parameters $m_0$ ($0-500\;GeV/c^2$) and $m_{1/2}$ ($50-200  \;GeV/c^2$).  
The PYTHIA~\cite{PYTHIA} Monte Carlo program with the ISASUGRA implementation is used to predict the SUSY spectrum at the TeV scale.  
Light flavor $\squ$ masses are considered degenerate, while 2-to-2 processes involving stop ($\sto$) and sbottom ($\sbo$) production are excluded to avoid strong theoretical dependence on the mixing in the third generation.
Signal samples are normalized to the next-to-leading-order (NLO) cross section calculated with PROSPINO~\cite{PROSPINO}.
The SM backgrounds are dominated by QCD multi-jets processes where $\MET$ comes from partially reconstructed jets in the final state.  
Other sources of backgrounds are Z and W production in association with jets, top and diboson production where $\MET$ is mostly due to the presence of neutrinos or muons in the final state.  
The estimation of both signal and background yields relies on MC techniques.  
The background model is tested in different control regions defined to enhance a particular background contribution.  
A good agreement is always observed between data and SM prediction in all the control regions.

\begin{figure}[h!]
\begin{center}
\includegraphics[width=8cm]{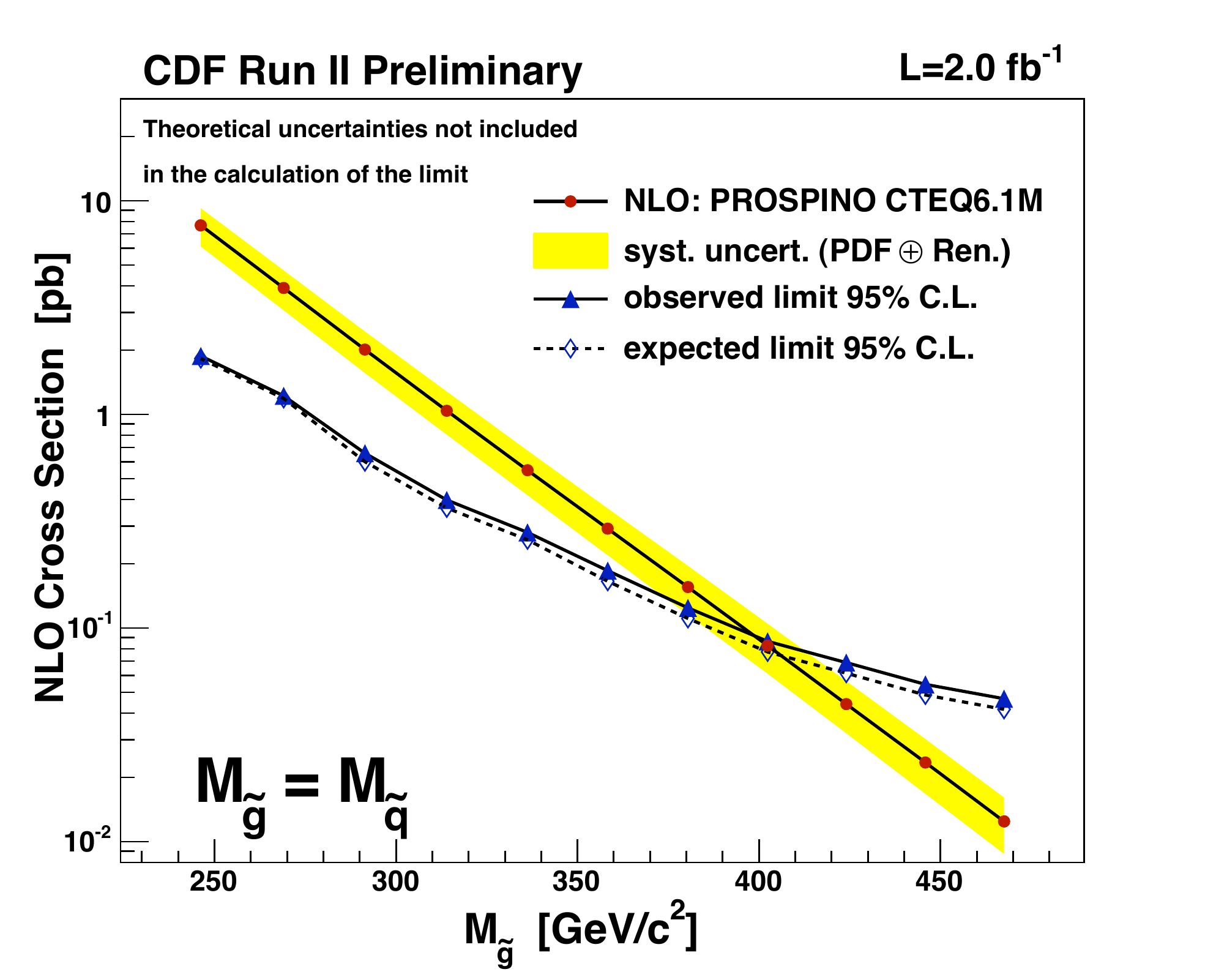}
\includegraphics[width=8cm]{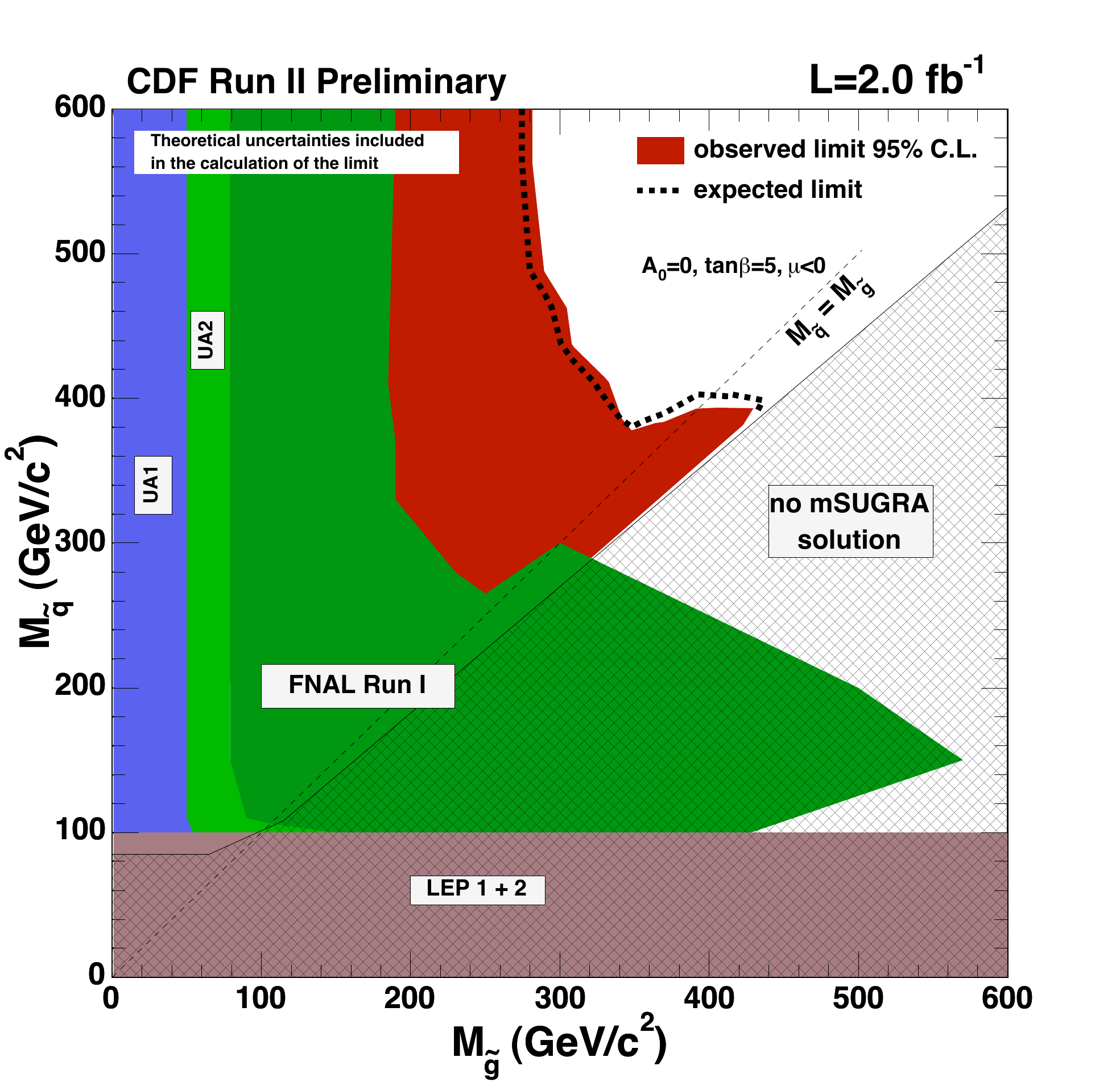}
\end{center}
\caption{\protect\footnotesize
\emph{left}: Exclusion limits on gluino and squark production cross section when $m_{\tilde{g}} = m_{\tilde{q}}$.
\emph{right}: Observed (red region) and expected (dashed-line) 95\% CL exclusion limits on gluino/squark production in the $m_{\tilde{g}}-m_{\tilde{q}}$ plane.  Regions excluded by previous experiments are shown, together with no-mSUGRA solution region.
\label{inclusive_limit}}
\end{figure}

For best sensitivity across the $m_{\tilde{g}}-m_{\tilde{q}}$ plane, three different analyses are developed, requiring at least 2, 3 or 4 jets in the final state. 
The rejection of distinct SM backgrounds relies on specific cuts on kinematic variables including the transverse energy of the leading jets, the $\MET$ of the event, and the angle between the $\MET$ and the direction of the jets.
Different selection criteria are defined for each final state to enhance the signal 
over background separation in different regions of the mSUGRA phase space.
A detailed study of the systematic uncertainties is carried out for each final state: the dominant systematic on the signal efficiencies and SM background yields comes from the $3\%$ uncertainty on the jet energy scale (JES).  
This translates into a $15\%$ and $30\%$ uncertainty on the signal and background estimation, respectively.
Other systematic uncertainties are related to the modeling of the initial- and final-state radiation (ISR/FSR) and the choice of Renormalization Scale and PDFs.  
Finally, a $6\%$ uncertainty on the integrated luminosity is also included.
The number of expected SM events is $48\pm17$ for the 4-jets analysis, $37\pm12$ for 3-jets and $16\pm5$ for 2-jets.
The observed events in $2\;fb^{-1}$ of data are 45, 38 and 18 respectively.
No significant deviation from the SM prediction is detected and the results are translated into exclusion limits on gluino and squark production.
The limits are computed with a Bayesian approach at $95\%$ confidence level (CL) and including all the systematics and their correlations in
the calculation.
Depending on gluino and squark masses, cross sections in the range $1\;pb$ to $0.1\;pb$ are excluded (see Fig.~\ref{inclusive_limit} left).  
In an mSUGRA scenario for which $m_{\tilde{g}} \sim m_{\tilde{q}}$ this analysis excludes masses up to $392\;GeV/c^2$.  
Squark masses below $280\;GeV$ are excluded in any case (see Fig.~\ref{inclusive_limit} right).

\section{SEARCH FOR SBOTTOM PRODUCTION FROM GLUINO DECAY}

Under the hypothesis $m_{\sbo} < m_{\glu} < m_{\squ}$, where $m_{\squ}$ is the mass of the light flavor squarks, the gluino pair production $p\overline{p}\rightarrow{}\glu\glu$ is one of the dominant SUSY processes at Tevatron.  
In the framework of the Minimal Supersymmetric Extension of the Standard Model (MSSM), the assumption $m_{t},m_{\cha} > m_{\sbo} > m_{\neu}$ leads to a branching ratio of $100\%$ for the $\sbo\rightarrow{}b\neu$ decay.  
In this configuration and with the conservation of the R-Parity, the sbottom production through gluino decays has a very distinctive signature with 4 $b$-jets and large missing transverse energy ($\MET$) due to neutralinos in the final state.

The present analysis~\cite{SBO} is based on $2.5fb^{-1}$ of data collected with the CDF detector during Run II.
The event selection follows the strategy developed for the inclusive analysis requiring a minimum $\MET$ per event and at least two jets in the final state.
To enhance the heavy flavor content, at least one of the two leading jets is required to be $b$-tagged.
The spectrum of MSSM events is predicted with PYTHIA while PROSPINO is employed to calculate the NLO cross section.
The gluino and the sbottom masses are varied from $240\;GeV/c^{2}$ to $400\;GeV/c^{2}$, and from $150\;GeV/c^{2}$ to $350\;GeV/c^{2}$, respectively.
The neutralino mass is fixed to $m_{\neu}=60\;GeV/c^{2}$ and the squark mass to $m_{\squ}=500\;GeV/c^{2}$.
The estimation of the dominant background contributions from QCD multi-jet production and light flavor jets mis-tagged as $b$-quark comes directly from data.
The remaining SM backgrounds, including the $t\overline{t}$ production, $Z$ and $W$ production in association with jets, single top and diboson production, are modeled using PYTHIA.
The SM prediction is tested in background dominated control regions in which signal contribution is negligible.
Good agreement between data and SM is always observed.
Different sources of systematic uncertainties are studied. 
The primary contributions come from the JES (up to $25\%$ on signal and background) and the \textit{mistags} estimation (around $5\%$).  

\begin{figure}[h!]
\begin{center}
\includegraphics[width=8cm]{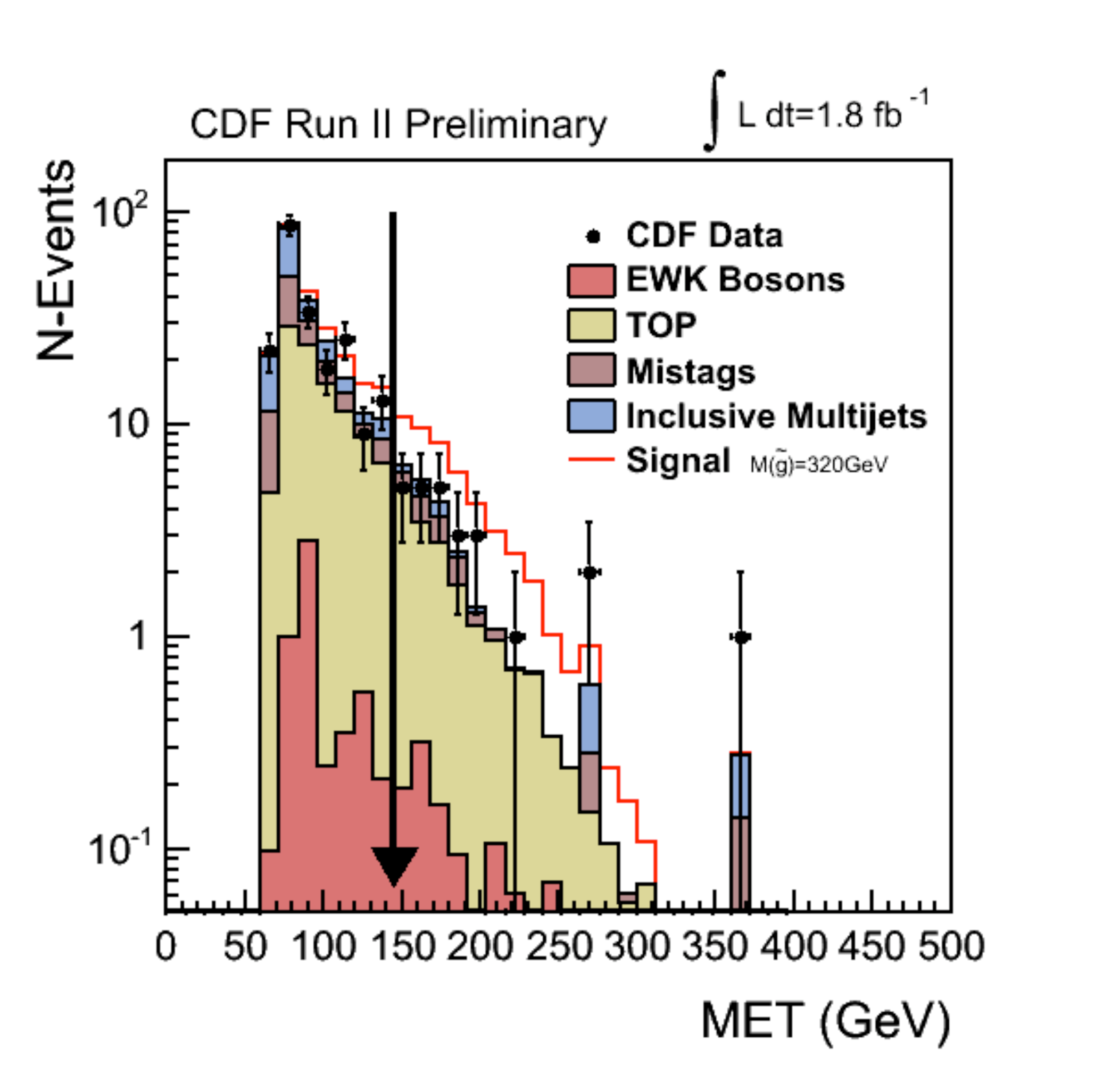}
\includegraphics[width=8cm]{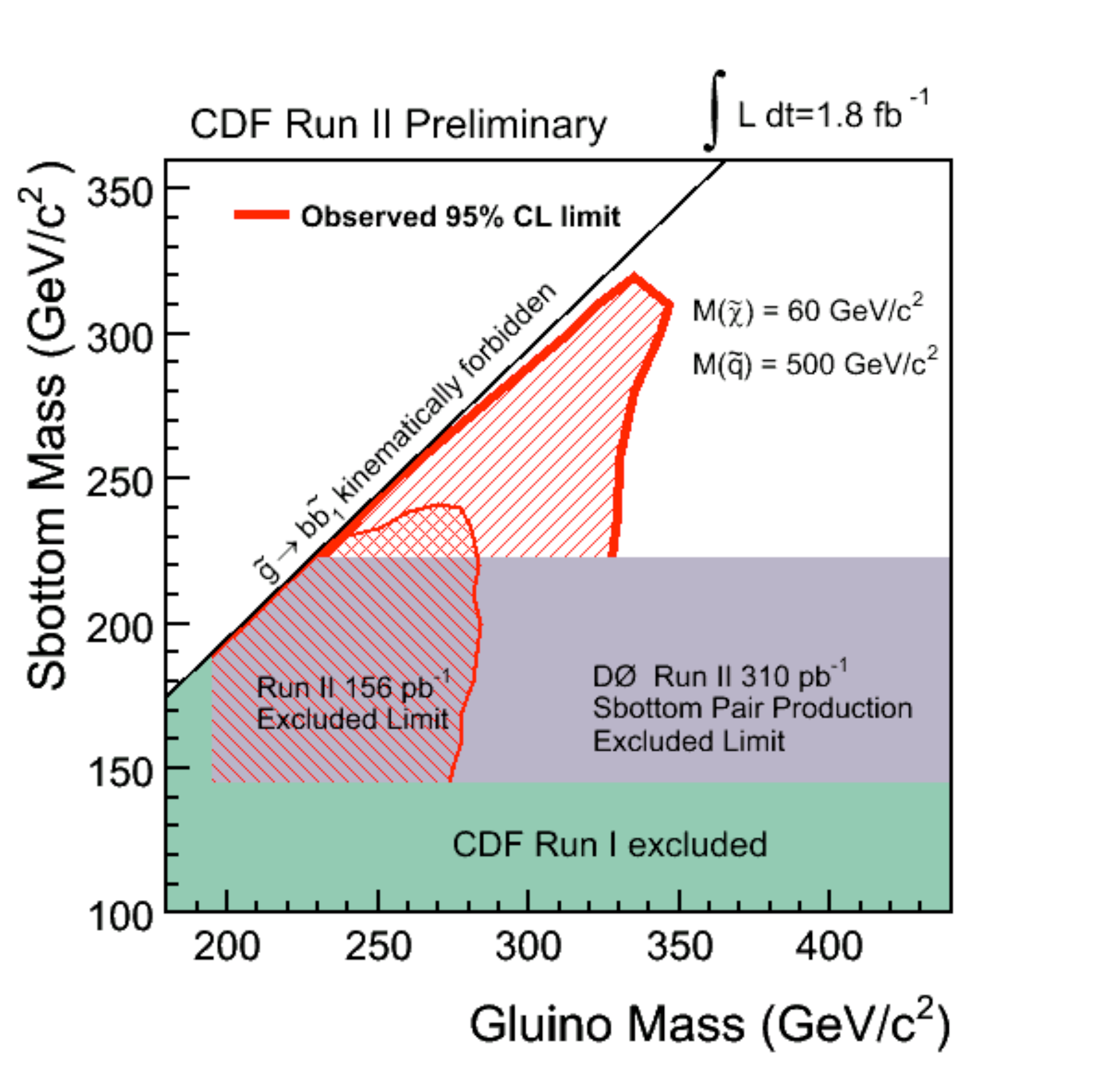}
\caption{\protect\footnotesize
\emph{left}: $\MET$ distribution in the large $\Delta{}m$ signal region.  Data are compared to expected SM and signal yields.
\emph{right}: Expected and Observed exclusion limits at $95\%$ CL on $m_{\sbo} - m_{\glu}$ plane.  
Limits from previous version of the analysis performed by CDF and D$\oslash$ during Run I and Run II\cite{SBOcdfII,SBOd0II} are also shown.
\label{sbottom_limit}}
\end{center}
\end{figure}

Final kinematic cuts on the $\MET$ of the event and the energies of the two leading jets are optimized to have the best sensitivity in two different areas of the SUSY phase space defined according to the value of $\Delta{}m = |m_{\glu}-m_{\sbo}|$. 
In the small $\Delta{}m \simeq 20\;GeV/c^{2}$ signal region 19 events are observed with a SM expectation of $22.0\pm{}3.6$.  In the large $\Delta{}m \simeq 80\;GeV/c^{2}$ signal region the expected SM yield is $22.7\pm{}4.6$ against 25 events observed in the data.
Since no significant deviation from the SM prediction was observed, the results are translated into an exclusion limit on sbottom production process.  
Limit is computed with a Bayesian approach at $95\%$ CL and including all the systematic uncertainties and their correlations in the calculation.
Cross sections down to $40\;fb$ are excluded.
Results are also translated into the exclusion limit on gluino and squark masses shown in Figure~\ref{sbottom_limit}: sbottom masses up to $320\;GeV$ are excluded for $m_{glu}\simeq{}335\;GeV$.

\end{document}